\documentclass[final]{elsart5p}
\usepackage{epsfig}
\usepackage{amssymb}
\begin{document}
\begin{frontmatter}
\title{Stephani-Schutz quantum cosmology}
\author{P. Pedram,}\author{ S.
Jalalzadeh\ead{s-jalalzadeh@sbu.ac.ir}\corauthref{cor1}}
\corauth[cor1]{Corresponding author}
\author{, and S. S. Gousheh}
\address{Department of Physics, Shahid Beheshti University,
Evin, Tehran 19839, Iran}

\begin{abstract}
We study the Stephani quantum cosmological model in the presence of a
cosmological constant in radiation dominated Universe. In the
present work the Schutz's variational formalism which recovers the
notion of time is applied. This gives rise to Wheeler-DeWitt equations which can be cast in the form of
Schr\"odinger equations for the scale factor. We find their
eigenvalues and eigenfunctions by using the Spectral Method. Then we use the eigenfunctions in order to construct wave packets and 
evaluate the time-dependent expectation value of the scale factor,
which is found to oscillate between non-zero finite maximum and minimum
values. Since the expectation value of the scale factor never tends
to the singular point, we have an initial indication that this model
may not have singularities at the quantum level.
\end{abstract}
\end{frontmatter}

\section{Introduction}
In recent years observations show that the expansion of the Universe
is accelerating  in the present epoch \cite{1} contrary to
Friedmann-Robertson-Walker (FRW) cosmological models, with
non-relativistic matter and radiation. Some different physical
scenarios using exotic form of matter have been suggested to resolve
this problem \cite{2,3,4,5,6,chap}. In fact the presence of exotic
matter is not necessary to drive an accelerated expansion. Instead
we can relax the assumption of the homogeneity of space, leaving the
isotropy with respect to one point. The most general class of
non-static, perfect fluid solutions of Einstein's equations that are
conformally flat is known as the ``Stephani Universe''
\cite{1-1,2-1}. This model can be embedded in a five-dimensional
flat pseudo-Euclidean space, which is not expansion-free and has
non-vanishing density \cite{9,1-1,3-3}. In general, it has no
symmetry at all, although its three dimensional spatial sections are
homogeneous and isotropic \cite{12}. The spherically symmetric
Stephani Universes and some of their subcases have been examined in
numerous papers \cite{2-1}. So it may be important to study the
quantum behavior of this model.

The notion of time can be recovered in some cases of quantum
cosmology, for example when gravity is coupled to a perfect fluid
\cite{14-2,R22-4,R22-6}. This kind of systems are often  studied as follows \cite{chap,R22-4,R22-6}. First one uses the Schutz's
formalism for the description of the perfect fluid \cite{11-2,12-2},
second  one selects the dynamical variable of perfect fluid as the
reference time. Finally, one uses canonical quantization to obtain
the Wheeler-Dewitt (WD) equation in minisuperspace, which is a
Schr\"odinger-like equation \cite{14-2}. After solving the equation,
one can construct wave packets from the resulting modes. The wave packets can be used to compute the time-dependent behavior of
the scale factor. If the selected time variable results in a close correspondence
between  the expectation value of the scale factor and  the
classical prediction (prediction of General Relativity) for long  enough
time, the selected time variable can be considered  as acceptable. This approach has been extensively employed
in the literature, indicating in general the suppression of the
initial singularity
\cite{chap,14-2,R23-10,nivaldo-2,R22-4,R23-13,R22-6}.

In the present paper, we use the formalism of quantum cosmology in
order to quantize the Stephani cosmological model in Schutz's formalism
\cite{11-2,12-2} and find WD equation in minisuperspace. In the
Schutz's variational formalism the wave function depends on the
scale factor $a$, and on the canonical variable associated to the
fluid, which  plays the role of time $T$. Here, we describe matter
as a perfect fluid matter. Although, this is essentially
semiclassical from the start, it has the advantage of defining a
variable, connected with the matter degrees of freedom, which can
naturally be identified with time and leads to a well-defined
Hilbert space structure. Moreover, after the universe reaches the
dust dominated matter, the evolution towards an exponentially
expanding epoch involves a quantum mechanical transition associated
with some vestige component of the original wave function of the
Universe. In fact, the classical Universe on large scales is based
on a quantum mechanical background. Particularly, a rapidly
oscillating state with small amplitude, which is the cosmological
influence of a wave function vestige, would emerge in the dust
dominated epoch \cite{Buahmadi}.
\section{The Model}
The action for gravity plus perfect fluid in Schutz's formalism is
written as
\begin{eqnarray}\nonumber
\label{action} S &=& \frac{1}{2}\int_Md^4x\sqrt{-g}\, (R-2\Lambda) +
2\int_{\partial M}d^3x\sqrt{h}\, h_{ab}\, K^{ab} \nonumber\\ &+&
\int_Md^4x\sqrt{-g}\, p \quad ,
\end{eqnarray}
where $K^{ab}$ is the extrinsic curvature, $\Lambda$ is the
cosmological constant, and $h_{ab}$ is the induced metric over the
three-dimensional spatial hypersurface, which is the boundary
$\partial M$ of the four dimensional manifold $M$ in units where
$8\pi G=1$ \cite{7-2}. The last term of (\ref{action}) represents
the matter contribution to the total action, and $p$ is the
pressure. In Schutz's formalism \cite{11-2,12-2} the fluid's
four-velocity is expressed in terms of five potentials $\epsilon$,
$\zeta$, $\xi$, $\theta$ and $S$:
\begin{equation}
u_\nu = \frac{1}{\mu}(\epsilon_{,\nu} + \zeta\xi_{,\nu} + \theta
S_{,\nu}),
\end{equation}
where $\mu$ is the specific enthalpy, the variable $S$ is the
specific entropy, while the potentials $\zeta$ and $\xi$ are
connected with rotation and are absent in models of the FRW type.
The variables $\epsilon$ and $\theta$ have no clear physical
meaning. The four-velocity is subject to the normalization condition
\begin{equation}
u^\nu u_\nu = -1.
\end{equation}
The metric in spherically symmetric Stephani Universe
\cite{9,10,1-1,12,2-1,14} has the following form,
\begin{eqnarray}\nonumber
ds^2 = &-&\left[F(t)
\frac{a(t)}{V(r,t)}\frac{d}{dt}\left(\frac{V(r,t)}{a(r,t)}\right)
\right]^2dt^2\\ &+&
\frac{a^2(t)}{V^2(r,t)}\left(dr^2+r^2d\Omega^2\right),\label{metric}
\end{eqnarray}
where the functions $V(r,t)$ and $F(t)$ are defined as
\begin{eqnarray}
  V(r,t) &=& 1+\frac{1}{4}k(t)r^2, \\
  F(t) &=& \frac{a(t)}{\sqrt{C^2(t)a^2(t)-k(t)}}.
\end{eqnarray}
Using the line element (\ref{metric}) and the Einstein's equation,
one can easily show the functions $C(t)$, $k(t)$ and $a(t)$ are not
all independent, but are related to each other with the following
expressions
\begin{eqnarray}
  \rho(t) &=& \frac{3C^2(t)+\Lambda}{8\pi G},\\
  p(r,t) &=& \frac{1}{8\pi G}[2C(t)\dot C(t)\frac{V(r,t)/a(t)}{(V(r,t)/a(t)\dot)}\nonumber \\
  &-& 3C^2(t)-\Lambda],
\end{eqnarray}
where an overdot denotes a derivative with respect to $t$. Note that
in the spherically symmetric Stephani models and the given
coordinate system, the energy density $\rho(t)$ is uniform, while
the pressure $p(r,t)$ is not and depends on the distance from the
symmetry center placed at $r = 0$. This is the reason why in such
models the barotropic equation of state (i.e. of the form $p =
p(\rho)$) does not exist. If, however, we assume some relations
between $\rho(t)$ and $p(r,t)$, this could allow us to eliminate one
of the unknown functions, e.g. $C(t)$. Hence we are left with two
unknown functions $k(t)$ and $a(t)$. The first one $k(t)$ plays the
role of a spatial curvature index, while the second one $a(t)$ is
the Stephani version of the FRW scale factor.

Now we consider an observer placed at the symmetry center of the
spherically symmetric Stephani Universe. All of our physical
assumptions will concern the neighborhood $r\approx0$. First of all
we assume that locally, matter filling up the Universe fulfils a
barotropic equation of state of the standard form
\begin{equation}\label{eqbarotropic}
 p(r\approx0, t) =\alpha \rho (t).
\end{equation}
By substituting the Stephani metric (\ref{metric}) in the
action (\ref{action}) and choosing a curvature function $k(t)$
in the form \cite{Godlowski}
\begin{equation}\label{k(t)}
k(t)=\beta a^{\gamma}(t),
\end{equation}
and after some thermodynamical considerations \cite{14-2}, the final
reduced effective action near $r\approx0$, takes the form
\begin{eqnarray}\nonumber
S &= &\int dt\biggr[-3\frac{\dot a^2a}{N} -\Lambda N a^3+ 3\beta
Na^{1+\gamma}\\ &+& N^{-1/\alpha} a^3\frac{\alpha}{(\alpha +
1)^{1/\alpha + 1}}(\dot\epsilon + \theta\dot S)^{1/\alpha +
1}\exp\biggr(- \frac{S}{\alpha}\biggl) \biggl],
\end{eqnarray}
where $N=F(t)a(t)\frac{d}{dt}\left(\frac{1}{a(t)}\right)$. The
reduced action may be further simplified by canonical methods
\cite{14-2} to the super-Hamiltonian
\begin{equation}
{\cal H} = - \frac{p_a^2}{12a}+\Lambda a^3 -3\beta a^{1+\gamma}
+\frac{
 p_\epsilon^{\alpha + 1}e^S}{a^{3\alpha}},
\end{equation}
where $p_a= -6{\dot aa}/{N}$ and $p_\epsilon = -\rho_0 u^0 Na^3$,
$\rho_0$ being the rest mass density of the fluid. Using the
canonical transformations
\begin{eqnarray}
T&=&p_Se^{-S}p_\epsilon^{-(\alpha + 1)},\quad \quad  \quad p_T =
p_\epsilon^{\alpha + 1}e^S  , \nonumber\\
\bar\epsilon &=& \epsilon - (\alpha + 1)\frac{p_S}{p_\epsilon},
\quad \quad \quad \bar p_\epsilon = p_\epsilon ,
\end{eqnarray}
which are the generalization of the ones used in Ref.~\cite{14-2},
the super-Hamiltonian takes the form
\begin{equation}
{\cal H} = - \frac{p_a^2}{12a} +\Lambda a^3- 3\beta a^{1+\gamma} +
\frac{p_T}{a^{3\alpha}} \,\, ,\label{EqHamiltonian}
\end{equation}
where the momentum $p_T$ is the only remaining canonical variable
associated with matter and appears linearly in the
super-Hamiltonian.

The classical dynamics is governed by the Hamilton equations,
derived from Eq. (\ref{EqHamiltonian}) and Poisson brackets, namely
\begin{equation}
\left\{
\begin{array}{llllll}
\dot{a} =&\{a,N{\cal H}\}=-\frac{\displaystyle Np_{a}}{\displaystyle 6a}\, ,\\
 & \\
\dot{p_{a}} =&\{p_{a},N{\cal H}\}=- \frac{N}{12a^2}p_a^2+3N(1+\gamma)\beta a^{\gamma}  \\
& \\
&-3N\Lambda a^2+\frac{3N\alpha}{a^{1+3\alpha}}p_T \, ,\\
& \\
\dot{T} =&\{T,N{\cal H}\}=Na^{-3\alpha}\, ,\\
 & \\
\dot{p_{T}} =&\{p_{T},N{\cal H}\}=0\, .\\
& \\
\end{array}
\right. \label{4}
\end{equation}
We also have the constraint equation ${\cal H} = 0$. Choosing the
gauge $N=a^{3\alpha}$, we have the following equations for the
system
\begin{eqnarray}\label{eqm1}
T&=&t,\\\nonumber \ddot{a}&=&(3\alpha-\frac{1}{2})\frac{\dot
a^2}{a}-\frac{1}{2}(1+\gamma)\beta
a^{6\alpha+\gamma-1}\\\label{eqm2} &+&\frac{1}{2}\Lambda
a^{6\alpha+1} -\frac{\alpha}{2}a^{3\alpha-2} p_T,\\\label{eqm3}
0&=&-\frac{3\dot a^2}{a^{6\alpha-1}}-3\beta a^{\gamma+1}+\Lambda
a^3+\frac{p_T}{a^{3\alpha}}.
\end{eqnarray}
Note that the classical equations for the case $\gamma=+1$, in
Ref.~\cite{Stelmach}, correspond with choosing $\Lambda=0$
and $N=1$. In this case ($\Lambda=0$, $N=1$) the constraint equation ${\cal H} = 0$
reduces to
\begin{eqnarray}
-3a\dot a^2-3\beta a^{2}+a^{-3\alpha}p_T=0,
\end{eqnarray}
or
\begin{eqnarray}
\left(\frac{da(t)}{dt}\right)^2+\beta
a(t)=\frac{p_T}{3a^{3\alpha+1}(t)}.
\end{eqnarray}

Imposing the standard quantization conditions on the canonical
variables ($p_a \rightarrow -i\frac{\displaystyle
\partial}{\displaystyle\partial a}$,  $p_T \rightarrow
-i\frac{\displaystyle\partial}{\displaystyle\partial T}$) and
demanding that the super-Hamiltonian operator (\ref{EqHamiltonian})
annihilate the wave function, we are led to the following WD
equation in minisuperspace ($\hbar =1$)
\begin{equation}
\label{sle} \frac{\partial^2\Psi}{\partial a^2} +(12\Lambda a^4-
36\beta a^{2+\gamma})\Psi - i12a^{1 -
3\alpha}\frac{\partial\Psi}{\partial t} = 0.
\end{equation}
According to the equation (\ref{eqm1}) $T=t$, can be associated with
the time coordinate \cite{lemos1999,lemos1998}. Equation (\ref{sle})
takes the form of a Schr\"odinger equation $i\partial\Psi/\partial t
= {\hat H} \Psi$. As discussed in \cite{nivaldo-2,lemos1998}, in
order for the Hamiltonian operator ${\hat H}$ to be self-adjoint the
inner product of any two  wave functions $\Phi$ and $\Psi$ must take
the form
\begin{equation}\label{inner}
(\Phi,\Psi) = \int_0^\infty a^{1 - 3\alpha}\Phi^*\Psi da.
\end{equation}
Moreover,  the wave functions should satisfy the restrictive
boundary conditions which the simplest ones are
\cite{lemos1998,18-2}
\begin{equation}
\label{boundary} \Psi(0,t) = 0 \quad \mbox{or} \quad
\frac{\partial\Psi (a,t)}{\partial a}\bigg\vert_{a = 0} = 0.
\end{equation}
The WD equation (\ref{sle}) can  be solved by separation of
variables as follows,
\begin{equation}
\Psi(a,t) = e^{iEt}\psi(a), \label{11}
\end{equation}
where the scale factor dependent part of the wave function
$\psi(a)$ satisfies
\begin{equation}
\label{sle2} -\psi'' +( 36 \beta a^{2+\gamma}-12\Lambda a^4)\psi
=12Ea^{1 - 3\alpha}\psi,
\end{equation}
and the prime denotes derivative with respect to $a$. The
interesting feature of the Stephani model is that the spatial
curvature is time-dependent. The recent observational data shows
that our Universe is spatially flat. Moreover, negative powers in
equation (\ref{k(t)}) lead to the spatially flat Universe in the
present epoch.

We construct a general solution to the WD equation (\ref{sle}) by
taking linear combinations of the $\psi_{n}(a)$'s,
\begin{equation}
\Psi (a,t) = \sum_{n=0}^{m} C_{n}\psi_n(a) e^{iE_{n}t},
\label{wavepacket}
\end{equation}
where the coefficients $C_n(E_n)$ will be fixed later by choosing
appropriate boundary conditions. we can compute the expected value
for the scale factor $a$ for arbitrary wave packets,
following the {\it many worlds interpretation} of quantum mechanics
\cite{many,22-2}. We may write the expected value for the scale factor
$a$, with regards to the inner product (\ref{inner}) as
\begin{equation}
\left\langle a\right\rangle_t  =
\frac{\int_{0}^{\infty}a^{2-3\alpha}\,|\Psi (a,t)|^2 da}
{\int_{0}^{\infty}a^{1-3\alpha}\,|\Psi(a,t)|^2 da}.
\label{meanvalue}
\end{equation}

\section{The Spectral Method}
To solve the resulting WD equation (\ref{sle2}) we use Spectral
Method (SM) \cite{SM1,SM2,SM3,Pedram} for finding the bound states
of (\ref{sle2}). SM is simple, fast, accurate, robust and stable.
This method consists of first, choosing a finite domain for the
approximate solution denoted by $2L$, and second taking the solution as a
finite superposition of the Fourier basis functions in this domain
which satisfy the periodic boundary condition. By substituting the
expansion into the differential equation a matrix equation is
obtained. By minimizing the first eigenvalue of the resulting matrix with
respect to the value of spatial domain $2L$, the optimized basis functions can be found. The accurate energy eigenvalues correspond
to the eigenvalues of the resulting matrix with optimized basis
functions. We only examine the bound states of this problem, i.e.
the states which are the square integrable. The general equation
that we want to solve can be written in the form
\begin{equation}\label{eq8}
f_1(x)\frac{d^2\psi(x)}{dx^2}+f_2(x)\frac{d\psi(x)}{dx}+f_3(x)\psi(x)=\varepsilon\psi(x),
\end{equation}
Any complete orthonormal set can be used for the SM. We use the
Fourier series basis. That is, since we need to choose a finite
subspace of a countably infinite basis, we restrict ourselves to the
finite region $-L<x<L$. The value of $L$ is determined by requiring
that the sought-after eigenfunctions have compact support in this
domain subject to the aforementioned optimization. This means that we can expand the solution as
\begin{eqnarray}
\psi(x)=\sum_{i=1}^{2} \sum_{m=0}^{\infty} A_{m,i} \,\,\,
u_i\left(\frac{m \pi x}{L}\right), \label{eq10}
\end{eqnarray}
where
\begin{eqnarray}
\left\{
  \begin{array}{ll}
  u_1\left(\frac{m \pi
x}{L}\right)=\frac{1}{\sqrt{LR_m}} \sin\left(\frac{m \pi
x}{L}\right), &  \\
  u_2\left(\frac{m \pi
x}{L}\right)=\frac{1}{\sqrt{LR_m}}\cos\left(\frac{m \pi
x}{L}\right), & \\
  \end{array}\hspace{0cm}R_{m}=\left\{
  \begin{array}{ll}
    2, & \hspace{0cm}\hbox{m=0,} \\
    1, & \hspace {0cm}\hbox{otherwise.}
  \end{array}\right.
\right.
\end{eqnarray}
That is we assume periodic boundary condition. We can also make the
following expansions
\begin{eqnarray}
f_1(x)\frac{d^2\psi(x)}{dx^2}=\sum_{m,i} B_{m,i} \,\,\,
u_i\left(\frac{m \pi x}{L}\right),\label{eq11}
\end{eqnarray}
\begin{eqnarray}
f_2(x)\frac{d\psi(x)}{dx}=\sum_{m,i} C_{m,i} \,\,\, u_i\left(\frac{m
\pi x}{L}\right),\label{eq12}
\end{eqnarray}
\begin{eqnarray}
f_3(x) \psi(x)=\sum_{m,i} D_{m,i} \,\,\, u_i\left(\frac{m \pi
x}{L}\right),\label{eq13}
\end{eqnarray}
where $B_{m,i}$, $C_{m,i}$ and $D_{m,i}$ are coefficients that can
be determined once $f_1(x)$, $f_2(x)$ and $f_3(x)$ are specified. By
substituting and using the differential equation of the Fourier
basis we obtain
\begin{eqnarray}
\hspace{-0.5cm}\sum_{m,i}\hspace{-1mm}\bigg[B_{m,i}+C_{m,i}+D_{m,i}\bigg]u_i\hspace{-1mm}\left(\frac{m
\pi
x}{L}\right)=\hspace{-1mm}\varepsilon\hspace{-1mm}\sum_{m,i}A_{m,i}\,u_i\hspace{-1mm}\left(\frac{m
\pi x}{L}\right)\hspace{-1mm}.\label{eq14}
\end{eqnarray}
Because of the linear independence of $u_i(\frac{m \pi x}{L})$,
every term in the summation must satisfy
\begin{eqnarray}
B_{m,i}+C_{m,i}+D_{m,i}=\varepsilon\, A_{m,i}.\label{eq15}
\end{eqnarray}
It only remains to determine the matrices $B$, $C$ and $D$. Using
Eq.~(\ref{eq10}) and Eqs.\ (\ref{eq11},\ref{eq12},\ref{eq13}) we
have
\begin{eqnarray}
\hspace{-0.2cm}\sum_{m,i}\hspace{-1mm} B_{m,i} u_i\left(\frac{m \pi
x}{L}\right)=-\hspace{-1mm}\sum_{m,i} A_{m,i}\left(\frac{m \pi
}{L}\right)^2\hspace{-1mm} f_1(x) u_i\hspace{-1mm}\left(\frac{m \pi
x}{L}\right),
\end{eqnarray}
\begin{eqnarray}
\sum_{m,i} C_{m,i} u_i\left(\frac{m \pi
x}{L}\right)\,\,\,=\sum_{m,i} A_{m,i}f_2(x)
\frac{d}{dx}u_i\left(\frac{m \pi x}{L}\right),
\end{eqnarray}
\begin{eqnarray}
\sum_{m,i} D_{m,i} u_i\left(\frac{m \pi
x}{L}\right)\,\,\,=\sum_{m,i} A_{m,i}f_3(x) u_i\left(\frac{m \pi
x}{L}\right).
\end{eqnarray}
By multiplying both sides of the above equations by $u_{i'}(\frac{m'
\pi x}{L})$ and integrating over the $x$-space and using the
orthonormality condition of the basis functions, one finds
\begin{eqnarray}\nonumber
B_{m,i}&=&-\sum_{m',i'} A_{m',i'}\left(\frac{m \pi}{L}\right)^2
\int_{-L}^{L}\hspace{-0.2cm}u_i\left(\frac{m \pi x}{L}\right) f_1(x)
u_{i'}\left(\frac{m' \pi x}{L}\right)dx,\\
&=& \sum_{m',i'} b_{m,m',i,i'} A_{m',i'},\label{int1}
\end{eqnarray}
\begin{eqnarray}\nonumber
C_{m,i}&=&\sum_{m',i'} A_{m',i'}
\int_{-L}^{L}\hspace{-0.2cm}u_i\left(\frac{m \pi x}{L}\right) f_2(x)
\frac{d}{dx}u_{i'}\left(\frac{m' \pi x}{L}\right)dx,\\
&=& \sum_{m',i'} c_{m,m',i,i'} A_{m',i'},\label{int2}
\end{eqnarray}
\begin{eqnarray}\nonumber
D_{m,i}&=&\sum_{m',i'} A_{m',i'}
\int_{-L}^{L}\hspace{-0.2cm}u_i\left(\frac{m \pi x}{L}\right) f_3(x)
u_{i'}\left(\frac{m' \pi x}{L}\right)dx,\\
&=& \sum_{m',i'} d_{m,m',i,i'} A_{m',i'}.\label{int3}
\end{eqnarray}
Therefore we can rewrite Eq.\ (\ref{eq15}) as
\begin{eqnarray}
\sum_{m',i'}\hspace{-0.1cm}\bigg[b_{m,m',i,i'}+c_{m,m',i,i'}+d_{m,m',i,i'}\bigg]
A_{m',i'}=\varepsilon\, A_{m,i}.\label{eq16}
\end{eqnarray}
By selecting a finite subset of the basis functions, {\it e.g.}
choosing the first $2N$ which could be accomplished by letting the
index $m$ run from 1 to $N$ in the summations, equation (\ref{eq16})
can be written as
\begin{eqnarray}
D\, A=\varepsilon\, A, \label{eq17}
\end{eqnarray}
where $D$ is a square matrix with $(2N) \times (2N)$ elements. Its
elements can be obtained from Eq.~(\ref{eq16}). The eigenvalues and
eigenfunctions of the Schr\"{o}dinger equation are approximately
equal to the corresponding quantities of the matrix $D$. That is the
solution to this matrix equation simultaneously yields $2N$ sought
after eigenstates and eigenvalues.
\section{Results}
\begin{figure}
\centering
  \includegraphics[width=7cm]{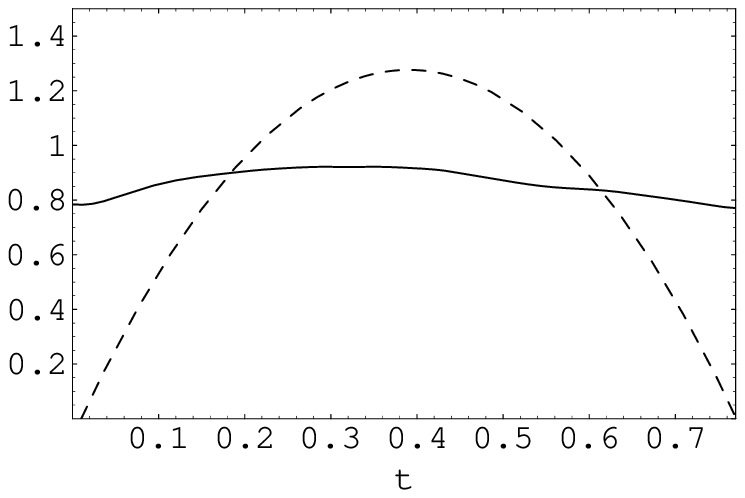}\\
  \caption{Classical behavior of the scale factor (dashed line) and the quantum mechanical expectation value of the scalar factor (solid line) for $\beta=1$,
  $\Lambda=1$, and $\gamma=4$ in radiation regime.}\label{fig1}
\end{figure}

\begin{figure}
\centering
  \includegraphics[width=7cm]{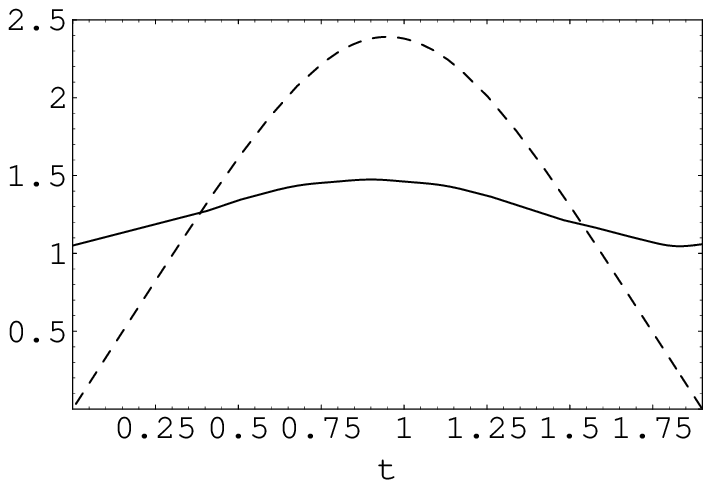}\\
  \caption{Classical behavior of the scale factor (dashed line) and the quantum mechanical expectation value of the scalar factor (solid line) for  $\beta=1$,
  $\Lambda=-1$, and $\gamma=-2$ in radiation regime.}\label{fig2}
\end{figure}

\begin{figure}
\centering
  \includegraphics[width=7cm]{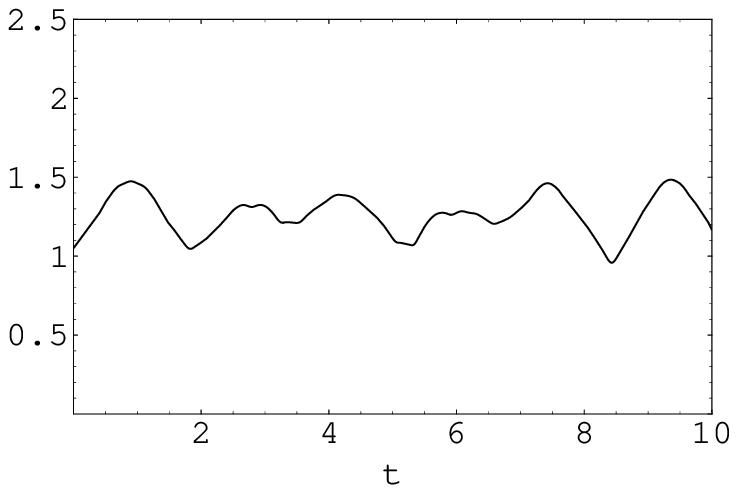}\\
  \caption{The expectation value of the scalar factor for  $\beta=1$,
  $\Lambda=-1$, and $\gamma=-2$ in radiation regime for a long period of time.}\label{fig3}
\end{figure}

\begin{figure}
\centering
  \includegraphics[width=7cm]{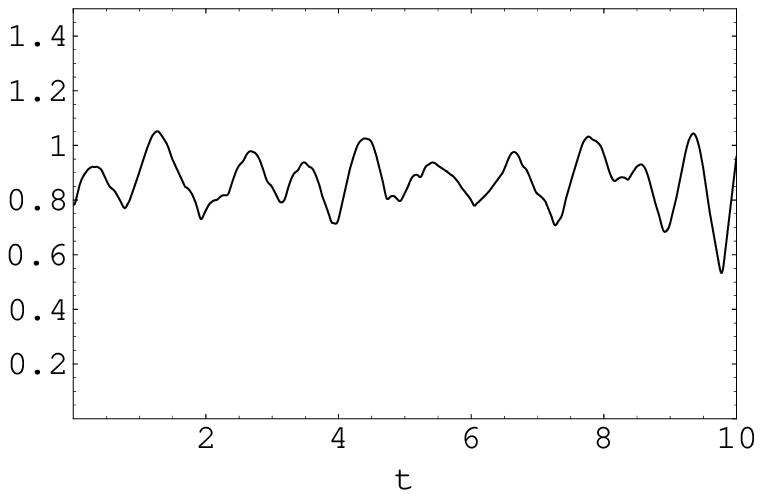}\\
  \caption{The expectation value of the scalar factor for $\beta=1$,
  $\Lambda=1$, and $\gamma=4$ in radiation regime for a long period of time.}\label{fig4}
\end{figure}

Since the Hamiltonian commutes with Parity operator, eigenstates
divide into odd and even categories. We choose the odd solutions
which satisfy the first boundary condition (\ref{boundary}). We are
free to choose the parameters $\beta$, $\gamma$, and $\Lambda$.
Unlike FRW models in which the bound states merely correspond to
negative cosmological constants \cite{18-2}, in this model we can
find the bound states with negative or positive cosmological
constant by choosing suitable value of $\gamma$. In radiation regime
($\alpha=1/3$), comparing equations (\ref{sle2}) and (\ref{eq8}) we
have
\begin{eqnarray}
f_1(x)&=&-1,\hspace{2.15cm} f_2(x)=0,\\ f_3(x)&=&36\beta
a^{\gamma+2}-12\Lambda a^4,\hspace{6mm}  \varepsilon=12E.
\end{eqnarray}
Here, we restrict ourselves to two cases: ($\beta=1, \gamma=4,
\Lambda=1$) and ($\beta=1, \gamma=-2, \Lambda=-1$). In radiation
dominated regime the expectation value of the scale factor can be
written as (\ref{meanvalue})
\begin{equation}
\left\langle a\right\rangle_t = \frac{\int_{0}^{\infty}a\,|\Psi
(a,t)|^2 da} {\int_{0}^{\infty}\,|\Psi(a,t)|^2 da}.
\end{equation}
Table \ref{Tab1} shows the first 20 odd energy eigenvalues
of (\ref{sle2}) for the two mentioned categories with 10 significant
digits. We can now construct the wave packets (\ref{wavepacket}) by
superimposing the resulting eigenfunctions (\ref{wavepacket}). Here,
we choose first 20 eigenfunctions ($m=20$) and, for simplicity
according to Ref. \cite{18-2}, select the coefficients equal to one
($C_n=1$) to incorporate equally all energy levels. As can be seen
from the classical equations of motion (\ref{eqm2},\ref{eqm3}), the
model has singularities at the classical level. At the quantum level
the probability density of finding the scale factor is
(\ref{meanvalue})
\begin{equation}
P(a,t)=a^{1-3\alpha}|\Psi(a,t)|^2.
\end{equation}
Since the probability density of finding the scale factor at $a=0$
in radiation regime is zero for odd solutions at all times
($\lim_{a\rightarrow0}|\Psi(a,t)|^2=0$), we have a initial
indication that these models may not have singularities at the
quantum level. Figures \ref{fig1} and \ref{fig2} show the behavior
of expectation value of the scale factor for two mentioned cases in
comparison with the classical behavior. Although, in classical case
the scale factors reach the zero axes, the expectation value of the
scale factors never tend to the singular point. This is depicted in
Figs.~(\ref{fig3},\ref{fig4}) which show the long time behavior of
the scale factors. This means that the big bang and big crunch
phenomena are absent at quantum level. Similar properties have been
discussed in \cite{14-2,18-2,chap} for FRW cosmological models.

\section{Conclusion}
In this work we have investigated perfect fluid Stephani quantum
cosmological model in the presence of cosmological constant. The use
of Schutz's formalism allows us to obtain a Schr\"odinger-like WD
equation in which the only remaining matter degree of freedom plays
the role of time. We have obtained eigenfunctions and therefore
acceptable wave packets have been constructed by appropriate linear
combination of these eigenfunctions. The time evolution of the
expectation value of the scale factor has been determined in the
spirit of the many-worlds interpretation of quantum cosmology. We
have shown that, contrary to the classical case, the expectation
values of the scale factor avoid singularity in the quantum case.
\begin{table}

 \centering
\begin{tabular}{|c|c|c|}
  \hline
   & $\beta=1, \gamma=4, \Lambda=1$&$\beta=1, \gamma=-2, \Lambda=-1$   \\ \hline
$E_{1}$&0.7404299830&3.724923306\\ \hline
$E_{2}$&2.716898545&5.221651006\\ \hline
$E_{3}$&5.467397344&7.051977995\\ \hline
$E_{4}$&8.816858136&9.123953661\\ \hline
$E_{5}$&12.67642129&11.39097635\\ \hline
$E_{6}$&16.98807771&13.82407714\\ \hline
$E_{7}$&21.71008635&16.40315278\\ \hline
$E_{8}$&26.81052795&19.11327742\\ \hline
$E_{9}$&32.26395981&21.94284703\\ \hline
$E_{10}$&38.04947945&24.88253248\\ \hline
$E_{11}$&44.14951426&27.92463946\\ \hline
$E_{12}$&50.54902037&31.06269279\\ \hline
$E_{13}$&57.23492813&34.29115371\\ \hline
$E_{14}$&64.19574440&37.60522022\\ \hline
$E_{15}$&71.42125871&41.00068165\\ \hline
$E_{16}$&78.90232098&44.47381007\\ \hline
$E_{17}$&86.63066966&48.02127735\\ \hline
$E_{18}$&94.59879650&51.64009065\\ \hline
$E_{19}$&102.7998386&55.32754141\\ \hline
$E_{20}$&111.2274909&59.08116446\\ \hline
\end{tabular}
\caption{The lowest calculated odd energy levels for two cases in
radiation dominated Universe.} \label{Tab1}
\end{table}


\end{document}